# Rapid droplet leads the Liquid-Infused Slippery Surfaces more slippery

Kun Li[1], Cunjing Lv[1,2,*] and Xi-Qiao Feng[1,2,*]

[1] *Institute of Biomechanics and Medical Engineering, Applied Mechanics Laboratory, Department of Engineering Mechanics, Tsinghua University, Beijing 100084, China*

[2] *Center for Nano and Micro Mechanics, Tsinghua University, Beijing 100084, China*

The introduction of lubricant between fluid and substrate endows the Liquid-Infused Slippery Surfaces with excellent wetting properties: low contact angle, various liquids repellency, ice-phobic and self-healing. Droplets moving on such surfaces have been widely demonstrated to obey a Landau–Levich–Derjaguin (LLD) friction. Here, we show that this power law is surprisingly decreased with the droplet accelerates: in the rapid droplet regime, the slippery surfaces seem more slippery than LLD friction. Combining experimental and numerical techniques, we find that the meniscus surrounding the droplet exhibits an incompletely developed state. The Incompletely Developed Meniscus (IDM) possesses shorter shear length and thicker shear thickness than the prediction of Bretherton model and therefore is responsible for the more slippery regime. With an extended Bretherton model, we not only provide an analytical description to the IDM behavior but also the friction when the Capillary Number of the moving droplet is larger than $5\times10^{-3}$.

The strong hysteresis between droplets and solids prevalent in daily life expects surfaces with extreme liquid-repellency [1-3]. The inherent interpretation of this hysteresis is based on chemical or physical defects of the solid surfaces [4]. In order to improve mobility, smoothing the substrates would be a straightforward strategy, either by perfecting the surfaces or by introducing lubricants. Compared to the perfection of surfaces, the introduction of lubricants, either gas (Leidenfrost [5], Superhydrophobic surfaces [6]) or another immiscible liquid [2, 7], seems more feasible and has been extensively studied. However, the methods to trap a stable layer of gas film between droplets and solids both have their flaws [5, 6, 8, 9]. Therefore, since the proposal of impregnating lubricant oil between fluids and substrates with physical or chemical effects [2, 7], inspired by *Nepenthes* pitcher plants, the Liquid-Infused Slippery Surfaces has attracted extensive attentions for its promising performances: capable of repelling various liquids [2, 7, 10], maintain low hysteresis [2, 3, 7, 11], self-healing [2, 12], ice-phobic [13-15] and high stability [16, 17].

Among these potentials, how the droplets shed down SLIS is one important fundamental problem that has caught wild attention and bear significant implications [12, 18]. The first investigation about this topic was realized by Smith et al., where they explored the viscous dissipation originates from the lubricant film beneath the droplet, wetting ridge around the droplet and the droplet [11]. They found that, with the lubricant more viscous than the droplet, the dissipation is dominated in the wetting ridge with a Stokes-type friction which leads to a linear relationship between the friction and droplet velocity [11]. However, subsequent studies revealed a power-law nature instead of linear formulation between the drag force and speed due to the speed-dependent meniscus with different methods [19]. For example, with a cantilever force sensor, Daniel et al. measured the drag force acting on a sliding droplet directly and determined a Landau-Levich-Derjaguin (LLD) formula for the dissipation force to the speed [20]. This finding was also proved by regarding the movement of droplets down an inclined Liquid-Infused Slippery Surface as a uniform motion, when the balance between the friction and the gravity takes place [3]. While this power-law model has been confirmed and refined in further researches [19, 21], researchers have also identified its limitations to the cases of small Capillary Number (Ca) and leave large-Ca cases unsolved [3, 22], which, according to our subsequent study, is equally crucial in practical applications.

In this letter, with experimental, numerical and theoretical techniques, we extend our study of droplet movement on Slippery Surfaces to large-Ca cases and verify a more slippery regime with smaller scaling than the small-Ca LLD friction. Our systematic experiments distinctly indicate that, in conjunction with the evolving dynamics of droplet motion, the characteristics of meniscus undergo discernible modifications as well. On the other hand, this variation in meniscus characteristics precisely reflects, in the further simplified numerical simulations, the incomplete development state of meniscus, which possesses shorter shear length and thicker shear thickness than the Bretherton model predicts and therefore accounts for the more slippery regime. Finally, based on the aforementioned analysis, with an extended Bretherton model, we not only provide a reasonable explanation for the alteration of meniscus but also a theoretical model for rapid droplet movement on Slippery Surfaces.

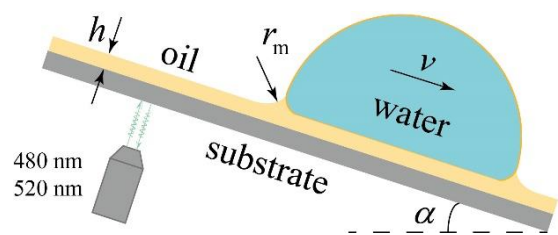

Fig. 1. Schematic of the experimental setup: a drop of deionized water of volume $V_0$ sliding down a Slippery Surface inclined at an angle $\alpha$ with velocity $v$. The thickness of film $h$ on the trailing trajectory of the droplet is measured using a modified fluorescence microscope (Nikon Ti2) and high-speed cameras with interference fringes. The dimension of meniscus $r_m$ is measured with zoom-in back-lighting photography.

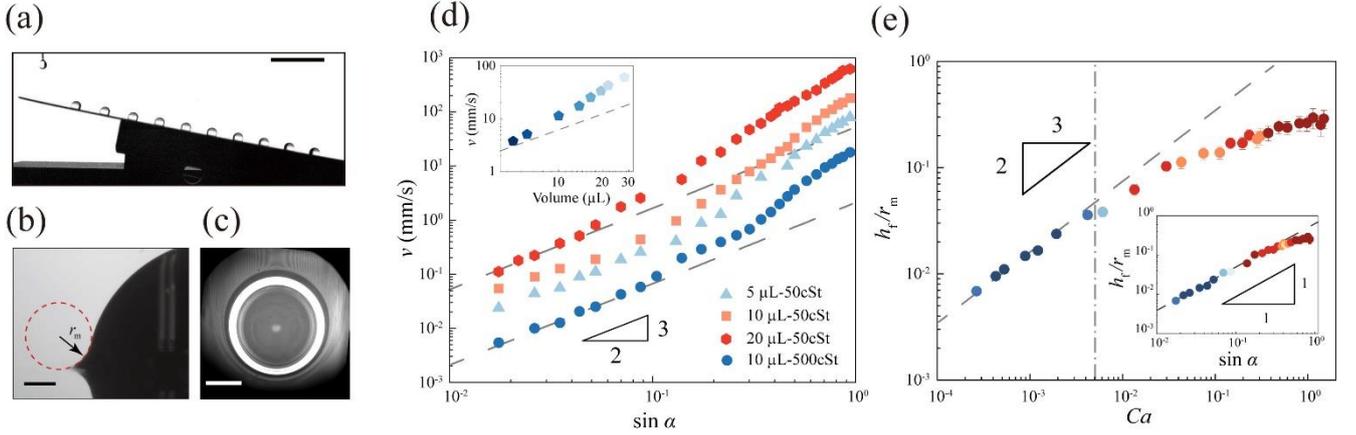

Fig. 2. Experimental results. (a) A typical chronophotograph with video frames for a 10 μL droplet sliding down an inclined Slippery Surface at 10°. The time interval between each pair of droplets is 2s and the scale bar is 10 mm. (b) A typical measurement of the meniscus dimension $r_m$. (c) Interferential fringes of 480nm for measurement of the film thickness $h$ at the tail of a 5 μL droplets sliding on a Slippery Surface spin-coated with 30 cSt silicon oil. (d) Dependence of droplet velocity $v$ on the inclined angle $\alpha$. Zoomed-in shows the effect of droplet volume $V_0$ on droplet velocity $v$ down a 15° inclined Slippery Surface. The black dashed line implies the LLD friction $v \sim (\sin\alpha)^{3/2}$. (e) The relationship between two meniscus characteristics, $h/r_m$, and the Capillary Number, $Ca = \eta_o v/\sigma_o$. The black dashed line implies the Bretherton prediction $h/r_m \sim Ca^{2/3}$.

Here, referring to the methodology of Keiser et al. by regarding the movement of droplets down an inclined Slippery Surface as a uniform motion [3, 19], we design the experimental setup as Fig. 1. Droplet velocities at different angles are extracted from back-lighting photographs with an Industrial Camera (IDS UI-3060 CP) after we drip deionized water onto the Slippery Surfaces. The Slippery Surfaces are placed on an angle-adjustable stage ($0° \leq \alpha \leq 70°$) and manufactured by spin-coating a uniform layer of silicone oil (DOW XIAMETER PMX-200, density $\rho_o$ = 963 kg/m³, surface tension $\sigma_o$ = 20.3 mN/m, viscosity $\nu_o = \eta_o/\rho_o$ = 30, 50, 100, 500 cSt) onto a smooth silicon wafer coated with Glaco (SOFT 99). The lubricant thickness has been altered with changing spin-coating speed but shows no influences on droplet average velocities. The volume of droplet (5μL ≤ V ≤ 20μL) can be adjusted independently. A typical chronophotograph with video frames for droplet movement is shown in Fig. 2 (a).

In order to characterize friction, we focus on the drop mobility, that is, the way speed depends on the gravitational driving force, which equals to the friction [3]. In Fig. 2 (d), the drop velocity $v$ is plotted as a function of the sine of the tilt angle $\alpha$ with various volume $V_0$ and various lubricant viscosity $\eta_o$. Here, we only consider scenarios where the lubricant viscosity significantly surpasses that of the droplet, meaning dissipation is expected to mainly take place in oil [3, 11]. The solid black line in Fig. 2 (d) is the LLD expectations based on former researches [3, 19, 20, 22], and only matches our experiments below a critical angle. It is evident that, for millimeter-scale droplets, this critical angle falls approximately within the range of 5 to 15 degree (varies with different volume and lubricant viscosity), which suggests that the LLD friction can only explain a very limited subset of droplets moving on tilted Slippery Surfaces and the significant implications of our study.

In order to uncover the underlying origins behind the variation in frictional behavior, a more detailed investigation of the dominant region of dissipation (the "foot" of oil [19], the wetting ridge [11, 22] or the meniscus [11, 22]), is imperative. Therefore, with a modified fluorescence microscope (Nikon Ti2) and two high-speed camera (Photron Fastcam Mini UX100 and Nova S20) with filters(480 nm and 520 nm), we conducted systematic measurements on the curvature radius $r_m$ of meniscus (as shown in Fig. 2 (b)) and the film thickness $h_f$ at the tail of a 20 μL droplets (as shown in Fig. 2 (c)) sliding on a Slippery Surface spin-coated with 50 cSt silicon oil on a 4 inches glass. The set up is shown as in Fig. 1. In Fig. 2 (e), since the variation of $h_f/r_m$ is not prominent in the $\sin\alpha$ coordinates (as shown in the inset figure of Fig. 2 (e)), we present the relationship between two characteristic variables, $h_f/r_m$, and the Capillary Number, $Ca = \eta_o v/\sigma_o$. Similarly, in comparison with predictions from the Bretherton model (black lines in Fig. 2 Fig. 1(e)), we discern that accompanying with the alterations in droplet motion behavior, there are also modifications in the configuration of meniscus. Therefore, it's reasonable to reconsider the viscosity dissipation originating from the meniscus.

The meniscus, as a universal fluid feature, has been observed in wide systems, including long bubbles moving in tubes[23, 24], plates or fibers pulled out of liquid [25-27], capillary rises on a solid wall [28, 29], and droplet spreading on a wet plate[30, 31]. Despite the theoretical prediction of meniscus in the limit of small capillary numbers has long been proposed based on lubrication approximation[23, 25, 26], the establishment of theoretical frameworks of more complex scenarios remain challengeable for their intricate two-dimensional characteristics [32, 33]. Similarly, in our experiments, the longitudinal dimension (scale of film thickness $h$) of the dynamic meniscus is no longer significantly smaller than its lateral dimension (scale of meniscus $r_m$), which means the lubrication approximation seems no longer valid [4]. Therefore, we turn to numerical simulations to rationalize the change of meniscus and friction.

Considering the intricate multiscale disparities between lubricant film thickness and droplet size in practical scenarios, our numerical simulations solely investigate the meniscus region and abstract it as a free surface problem [32, 34]. The study domain is sketched in Fig. 3 (a), where we consider a steady 2D flow of the lubricant with viscosity $\eta_o$ between two solid surfaces of length $L$ and $H$ distance apart. To avoid the occurrence of computationally unfavorable singularities, in the model, we replace the lubricant-water-air triple contact line with a fixed contact angle of 90° between lubricant-air interface and upper surface. We designate the left side of the solution domain as a *pressure outlet* and the right side as a *completely developed mass flow*, where the flow rate is tuned until the interface profile stays still [35]. Simultaneously, the lower boundary is shifted to the negative $x$ direction at the same velocity $v$ as the experimental droplet. We employ the Navier-Stokes equations to describe the lubricant flow and neglect the viscosity of air and gravitational effect as the scale of the meniscus (100 μm) is smaller than the Capillary Length ($l_c \sim \sqrt{\sigma_o/\rho_o g} \sim 1.46$ mm).

When conducting numerical simulations of the meniscus, a crucial consideration is how to set the dimension of the computational domain to match experimental conditions. Is it necessary to incorporate the measured meniscus dimension as a known parameter in the simulation? Therefore, we initiated a series of calculation with different computational domain sizes for the Ca = $10^{-3}$ case and extracted the trailing thickness $h$, as shown in Supplemental Material, Fig. S1 [36]. The linear correlation between the trailing thickness $h_f$ and the computational domain size $H$ indicates the domain size independence.

Leveraging our simulations, we investigated the detailed variations of the curved meniscus. In Fig. 3 (b), we plot side by side the profiles of meniscus under three distinct $Ca$ = 0.001, 0.01 and 0.1, along with the distributions of shear stress on the substrate, respectively. Together, they elucidate the evolution of the meniscus with $Ca$. From the distribution of shear stress, we find that, in each case, there is a concentrated region. This region is abstracted and designated as dynamic meniscus as only within this region the shear stress significantly affects overall resistance of the system [33]. Meanwhile, we refer to the length of this region as the Length of Dynamic Meniscus $\lambda$, signifying the characteristic viscous shear length [3]. Yet, their scales are qualitatively different.

For small Capillary Number ($Ca = 10^{-3}$), along the outlet-to-inlet direction, shear stress increases from nearly zero to a maximum value and then returns to nearly zero before the *x* position of contact line. In this scenario, the dynamic meniscus exists as the transition region between the wetting film and circular static meniscus, and its length has been discussed to scale with $\lambda \sim r_m Ca^{1/3}$ [25, 26]. This kind of meniscus exhibits a completely developed state.

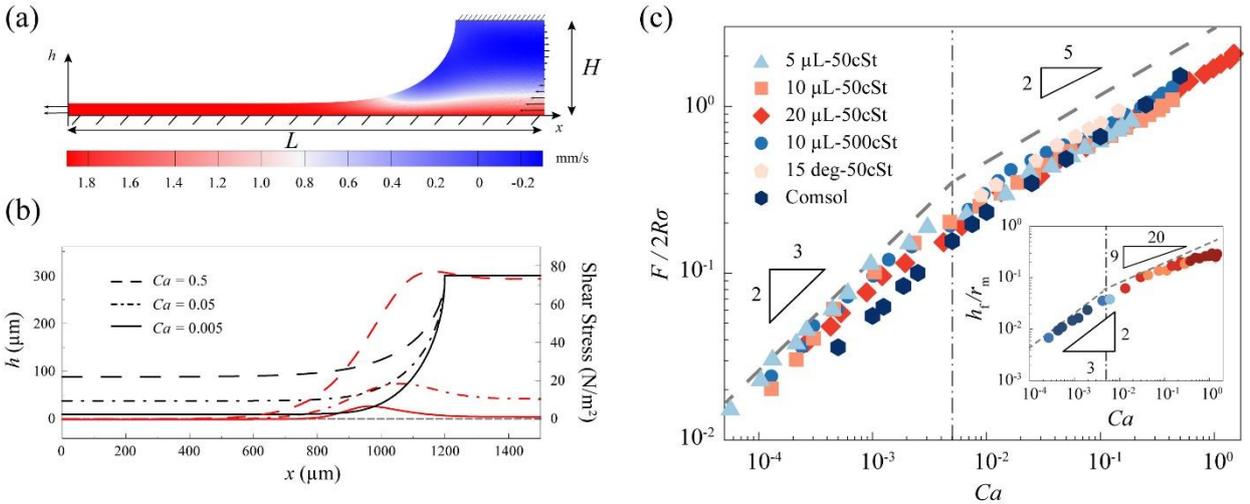

Fig. 3. (a) Numerical setup for the lubricant flow in meniscus, where the vectors and colors show the velocity field. (b) Profiles of interfaces (the black line) and shear stresses on the substrates (the red line) for Ca = 0.001, 0.01 and 0.1. The profiles and stresses are normalized with $H$ and largest shear stress, respectively. (c) Dependence of normalize friction on Capillary Number $Ca$. The black line is predicted by Bretherton, $\underline{F} \sim Ca^{2/3}$ while the orange line is the more slippery regime, $\underline{F} \sim Ca^{2/3}$. Zoom-in shows our empirical scaling function between meniscus features and Capillary Number $Ca$.

Conversely, for large Capillary Number ($Ca = 10^{-1}$), along the outlet-to-inlet direction, shear stress increases from nearly zero to a maximum value, but returns to a certain value at the *x* position of the contact line and keeps constant till the outlet. In this scenario, the dynamic meniscus is cut off at the *x* position of the contact line and the meniscus comprises only the wetting film and the dynamic meniscus, exhibiting an incompletely developed state. Given that in real-world scenarios, the meniscus does terminate at the oil-water-gas triple contact line, which has two effects on the viscous shear length $\lambda$ when accounting for viscous resistance: First, the characteristic dimension of dynamic meniscus, should be taken as the characteristic dimension of the meniscus, i.e., $r_m$. Second, "cutoff coefficient" should be introduced as the integral limit has changed from the entire Gaussian type shear stress to part of the entire Gaussian type shear stress, which is order of $Ca^{-0.15}$ as shown in Supplemental Material, Fig. S2 [36]. Therefore

$$\lambda \sim \begin{cases} r_m Ca^{1/3} & Ca < Ca_c \\ r_m Ca^{-0.15} & Ca > Ca_c \end{cases} \quad (1)$$

To elucidate the movement behavior of droplets on Slippery Surfaces, we employ the scaling law analysis to establish the dependent relationship between viscous resistance $F_\eta$, and the capillary number $Ca$. The typical velocity gradient in the meniscus scales as $v/h_f$, which yields a typical shear stress $\eta_o v / h_f$. Once integrated over the dynamic meniscus $\lambda$ and the droplet perimeter $2\pi R$, we get the total viscous friction

$$F_\eta \sim \frac{2\pi R \lambda \eta_o v}{h_f} \quad (2)$$

For small Capillary Number cases, the length of dynamic meniscus and film thickness is given as [3, 19, 20, 22, 25, 26, 33], $\lambda \sim r_m Ca^{1/3}$ and $h \sim r_m Ca^{2/3}$, which leads to a 2/3 power between the normalized friction $\bar{F}_\eta = F_\eta / 2\pi \sigma R$ and $Ca$, in accordance with the small-$Ca$ data as shown in Fig. 3 (c) and former researches [3, 19, 20, 22, 33]. For large Capillary Number cases, the length of dynamic meniscus is limited by the dimension of meniscus, $\lambda \sim r_m Ca^{-0.15}$. As for the wetting film thickness, for the moment, we utilize the experimental results without explanation, $h \sim r_m Ca^{0.45}$. Hence, we obtain the large Capillary Number regime, $\underline{F}_\eta \sim Ca^{0.4}$, as shown in Fig. 3 (c).

Until now, we can infer that the alteration in the friction of droplet moving on Slippery Surfaces with increase of capillary number origin from the changes in the meniscus, i.e., the meniscus gradually loses enough space to completely develop and incompletely developed meniscus possess shorter shear length $\lambda$ and thicker shear thickness $h_f$. Consequently, the rapid droplet leads the Slippery Surfaces more slippery. However, as we mentioned above, we provided the experimental shear thickness $h_f$ without explanation for large Capillary Number cases. Although the limitations of the "Lubrication Approximation" in describing cases with large capillary number cases have been mentioned previously, we think that with a small modification, the Bretherton model can be extended to large Capillary Number cases. Therefore, in the subsequent analysis, we will introduce the Extended Bretherton model to describe film thickness and provide a theoretical description of the droplet resistance on Slippery Surfaces.

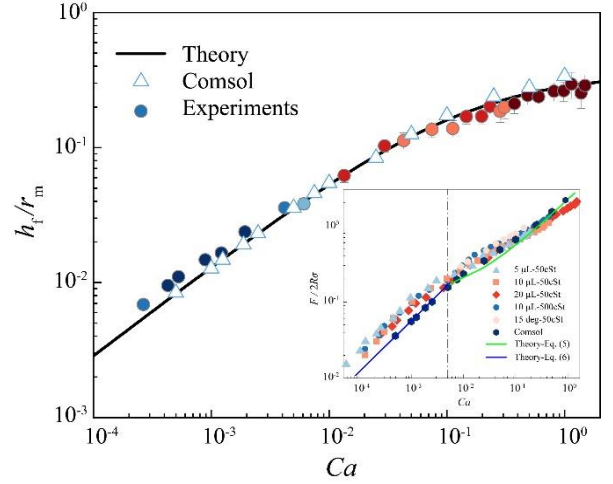

Fig. 4. Plot of the film thickness – meniscus dimension aspect ratio, $h_f/r_m$, as a function of Capillary Number, $Ca$. The plot contains experiments, (green squares), simulations (cyan hexagons) as well as the curve corresponding to the Extended Bretherton Model of Eq. (5) (red solid line). The green line in the embedded figure shows the comparison between the theoretical model of friction based on the Extended Bretherton Model and experiments for larger Capillary Number cases.

In the analysis of Bretherton on the motion of long bubbles in tubes [23], his focus was on the detail of "transition region", i.e., the dynamic meniscus. Based on the "Lubrication Approximation", Bretherton established the equation of motion of dynamic meniscus with three simplifications on the Navier-Stokes Equation: first, the velocity at every location is mainly in the direction of *x*, which simplifies the three-dimensional vector equation to a much friendlier scalar equation; second, because of the no-slip boundary condition on the wall, the velocity in the thin film was so slow that the nonlinear inertial term can be neglected compared to the second-order viscus term; third, likewise, due to the no-slip boundary condition on the wall and thin film geometry, the velocity gradient is primarily perpendicular to the flow, and hence decouples the driving force and the resistance force. In conjunction with the no-slip boundary condition on the wall, and the zero tangential stress condition and the normal stress condition at the interface, the equation of motion is integrated to a parabolic velocity profile, or a Poiseuille

profile. Finally, since the motion is steady, the flux conservation leads to the central equation

$$3Ca(h_f - h) = \frac{\partial^3 h}{\partial x^3} h^3 \quad (3)$$

where $h$ implies the profile of dynamic meniscus, which is a function of $x$, and is a constant. The non-dimensional formula of this equation is usually referred to as the Landau-Levich equation [37]. Although it does not have an analytical solution, there are asymptotic solutions both at the direction of wetting film and at the direction of contact line

$$h = \frac{A(3Ca)^{2/3}(x-x_0)^2}{2h_f} + Bh_f \quad (4)$$

This is actually a parabolic approximation of a sphere with curvature $\kappa_p = 1/r_p = A(3Ca)^{2/3}/h_f$, where $A$ and $B$ are constants. For a nonnegligible wetting film thickness compared to the asymptotic sphere radius, the dimension of meniscus scales with $r_m = Bh_f + r_p$, which leads

$$\frac{h_f}{r_m} = \frac{A(3Ca)^{2/3}}{1 + AB(3Ca)^{2/3}} \quad (5)$$

This particular expression is analogous to the one proposed By Bretherton in Ref. [23], except that we take into account the influence of wetting film on the characteristic dimension of meniscus. The constants $A$ and $B$ have been calculated numerically to be $A \approx 0.643$ and $B \approx 2.79$ in former studies [30]. When plotted in Fig. 4, our Extended Bretherton Model shows well agreement with the observed meniscus feature $h_f/r_m$ from small Capillary Number cases to large Capillary Number cases.

Further, by substituting Eq. (5) into Eq. (2), we get the theoretical expression on friction of rapid droplet moving on Slippery Surfaces

$$F_\eta \approx \frac{2\pi R \sigma Ca^{0.18}}{A}\left(1 + AB(3Ca)^{2/3}\right) \quad (6)$$

which shows great agreement with experiments as well in the embedded figure in Fig. 4. As for the approximate scaling law between $h_f/r_m$ and $F_\eta$ with respect to $Ca$, we can provide explanations together by considering the change of introduced term, $1+AB(3Ca)^{2/3}$. When $Ca$ change from $5\times10^{-3}$ to 0.5, this term change from 1.117 to 3.52, then we can approximate this term with an exponential form, $1+AB(3Ca)^{2/3} \approx Ca^{0.25}$, which leads to the approximate scaling laws $h_f/r_m \approx Ca^{0.42}$ and $F_\eta \approx Ca^{0.43}$.

Based on the analysis above, we can determine the critical Capillary Number from two perspectives. First, transition happens when the shear length for small-$Ca$ cases $\lambda \approx 5.1456 r_m Ca^{1/3}$ approaches meniscus dimension $r_m$, which means $5.1456 Ca^{1/3} \approx 1$ and derives the critical Capillary Number $Ca_c \approx 7.34\times10^{-3}$. Second, transition happens when the modification term of Extended Bretherton Model relative to the Bretherton Model, $1+AB(3Ca)^{2/3}$, starts to function, which means $AB(3Ca)^{2/3} \approx 0.1$ and leads to the critical Capillary Number $Ca_c \approx$ $4.38\times10^{-3}$. As shown in Fig. 2(c), Fig. 3(c) inset of Fig. 3(c) and inset of Fig. 4, the critical Capillary Number observed in our experiments, $Ca_c \approx 5\times10^{-3}$, exhibits a notable agreement with the critical Capillary Numbers determined from these two different perspectives.

In this Letter, we have presented our work on droplet moving on Lubricant-infused Slippery Surfaces, especially the rapid droplets. It was found the friction for rapid droplets does not behaves in the form of LLD power-law any more but a more slippery friction law. To explain this surprising tendency, we developed experimental and numerical techniques, and verified the incompletely developed state of meniscus for Large-$Ca$ cases. The incompletely developed meniscus was proved to possess shorter shear length and thicker shear thickness than the Bretherton model predicts, and therefore accounts for the a later more slippery regime. Finally, with an extended Bretherton model, we not only provide the theoretical description for meniscus thickness but also the friction for large-$Ca$ cases.

This work received financial support from the National Natural Science Foundation of China (No. XXX).